\documentclass[aps,prl,twocolumn,superscriptaddress,showpacs]{revtex4}
\usepackage{epsf}
\usepackage{amsmath,amssymb}
\usepackage{graphicx}
\usepackage{color}

\begin{document}

\title{How Well a Chaotic Quantum System\\ Can Retain Memory of Its
Initial State?}

\date{\today}

\affiliation{Budker Institute of Nuclear Physics, Novosibirsk,
Russia}
\author{Valentin V. Sokolov}
\affiliation{Budker Institute of Nuclear Physics, Novosibirsk,
Russia}
\affiliation{CNISM, CNR-INFM, and Center for Nonlinear
and Complex Systems, Universit\`a degli Studi dell'Insubria,
Via Valleggio 11, 22100 Como, Italy}
\author{Oleg V. Zhirov}
\affiliation{Budker Institute of Nuclear Physics, Novosibirsk,
Russia}

\begin{abstract}
In classical mechanics the local exponential instability effaces the
memory of initial conditions and leads to practical irreversibility.
In striking contrast, quantum mechanics appears to exhibit strong memory
of the initial state. We relate the latter fact to the low (at most linear)
rate with which the system's Wigner function gets during evolution more and
more complicated structure and establish existence of a critical strength
of external influence below which such a memory still survives.
\end{abstract}


\maketitle

Strong numerical evidence has been obtained~\cite{arrow} that
the quantum evolution of classically chaotic systems is quite
stable, in sharp contrast to the extreme sensitivity of
classical dynamics to initial conditions and weak external
perturbations. Being the very essence of classical dynamical
chaos, this sensitivity results in rapid loss of memory and
practical irreversibility of classical motion. Qualitatively,
this crucial difference is explained by a much simpler
structure of quantum states as compared to the extraordinary
complexity of random and unpredictable classical
trajectories~\cite{ford,alekseev}. In more rigorous terms,
chaotic classical systems are characterized by positive
\emph{algorithmic complexity} described by Lyapunov exponent.
Unfortunately, being formulated in such a way the concept of
complexity cannot be transferred, \emph{sic et simpliciter}, to
quantum mechanics, where the very notion of trajectory is
irrelevant and there is no quantum analogue to the Lyapunov
exponent. Therefore, at first glance, there exists no
quantitative measure of comparative complexity of classical and
quantum states of motion~\cite{prosencomplexity}.

However, individual classical trajectories are, in essence, of
minor interest if the motion is chaotic. They all are alike in
this case and rather behavior of manifolds of them carries
really valuable information. Therefore the methods of the phase
space and the Liouville form of the classical mechanics become
the most adequate. It is very important that, opposite to the
classical trajectories, the classical phase space distribution
and the Liouville equation have direct quantum analogs. Hence,
a comparison between classical and quantum dynamics can be made
by studying the evolutions in time of the classical and quantum
phase space distributions expressed in similar canonical
variables and both ruled by linear equations.

The paramount property of the classical dynamical chaos is the
exponentially fast structuring of the system's phase space on
finer and finer scales. On the contrary, degree of structuring
of the corresponding quantum "distribution" is restricted by
the quantization of the phase space. This makes Wigner function
relatively "simple" as compared to its classical counterpart.
The great advantage of the phase space approach is also that
one operates with distributions which can be presented in both
classical as well as quantum cases in identical action - angle
$(I, \theta)$ variables.

In practice, to explore stability or reversibility of motion
one computes \emph{fidelity}~\cite{peres,prosen} which
characterizes the weighted-mean distance between two
distributions evolving, for example, under two slightly
different Hamiltonians. In this paper, we exploit a somewhat
different aspect of the Peres fidelity. Namely, we use it to
establish a quantitative relation between the degree of loss of
memory on the one hand and complexity of the corresponding
phase space distribution on the other hand. In particular, we
show how \emph{the number ${\cal M}(t)$ of $\theta$-Fourier
harmonics} (see below Eqs. (\ref{FidRot}, \ref{tildeW_m})) can
be used as a suitable measure of complexity of a distribution
at a given time $t$. In the case of classical chaotic dynamics,
this number grows exponentially in time with the rate related
to the rate of local exponential instability~\cite{brumer}. At
the same time, this number increases only power-like if the
motion is regular. Thus the rate of growth of the number of
harmonics is, similar to the Lyapunov exponent, a
characteristic measure of classical complexity. In the
framework of the phase space approach, the number of harmonics
of the Wigner function appears, contrary to the Lyapunov
exponent, to be a relevant measure of complexity also in
quantum case. In what follows, we examine the time behavior of
this quantity and its relation to fidelity and reversibility
properties. A detailed derivation of the results reported in
this paper can be found in Ref.~\cite{harmonicslong}.

Let ${\hat H}\equiv H({\hat a}^{\dag},{\hat a};t)=H^{(0)}({\hat
n}={\hat a}^{\dag}{\hat a})+H^{(1)}({\hat a}^{\dag},{\hat
a};t)$ be the Hamiltonian of a generic nonlinear system with a
bounded below discrete energy spectrum $E_n^{(0)}\geq 0$, which
is driven by a time-dependent force of such a kind that the
classical motion exhibits a transition from integrable to
chaotic behavior when the strength of the driving force is
increased. Here ${\hat a}^{\dag}, {\hat a}$ are the bosonic,
creation-annihilation operators: $[{\hat a},{\hat
a}^{\dag}]=1\,$. We use the method of c-number $\alpha$-phase
space borrowed from the quantum optics (see for
example~\cite{Glauber63,Agarwal70}). It is basically built upon
the basis of the coherent states $|\alpha\rangle={\hat
D}\left(\frac{\alpha}{\sqrt\hbar}\right) |0\rangle$ obtained
from the ground state $|0\rangle$ of the unperturbed
Hamiltonian with the help of the unitary displacement operator
${\hat D}\left(\lambda\right) =\exp(\lambda\,{\hat
a}^{\dag}-\lambda^*{\hat a})$. Here $\alpha$ is a complex phase
space variable independent of the effective Planck's constant
$\hbar$.

The Wigner function $W$ in the $\alpha$-phase plane is defined
by the following Fourier transformation
\begin{equation}\label{Wfunc}
W(\alpha^*,\alpha;t)= \frac{1}{\pi^2\hbar}\int d^2\eta\,
e^{(\eta^*\frac{\alpha}{\sqrt\hbar}-
\eta\frac{\alpha^*}{\sqrt\hbar})} {\rm
Tr}\left[{\hat\rho(t)}\,{\hat D(\eta)}\right],
\end{equation}
where $\hat{\rho}$ is the density operator and the integration
runs over the complex $\eta$-plane. The Wigner function is
normalized to unity, $\int d^2\alpha W(\alpha^*,\alpha;t)=1$ and
is real though, in general, not positive definite. It satisfies
the evolution equation
\begin{equation}\label{Leq}
i\frac{\partial}{\partial t}\,W(\alpha^*,\alpha;t)= {\cal\hat
L}_q\,W(\alpha^*,\alpha;t),
\end{equation}
with the Hermitian ``quantum Liouville operator'' ${\cal\hat L}_q$.
This equation reduces in the case $\hbar=0$ to the classical
Liouville equation with respect to the canonical pair $\alpha,
i\alpha^*$ with the classical Hamiltonian function being given by
the diagonal matrix elements $H_c(\alpha^*,\alpha;t)=\langle\alpha|
{\hat H}^{(N)}({\hat a}^{\dag},{\hat a})|\alpha\rangle$ of the
normal form ${\hat H}^{(N)}$ of the quantum Hamiltonian operator.
In other words, this function is obtained from the quantum
Hamiltonian by substituting ${\hat a}\rightarrow
\alpha/\sqrt{\hbar}\,,\,\,{\hat a}^{\dag}\rightarrow\alpha^*/
\sqrt{\hbar}\,.$

We define the harmonic's amplitudes $W_m(I)$ as the Fourier
components of the Wigner function with respect to the angle
variable $\theta$ introduced by the canonical transformation
$\alpha=\sqrt{I}\,e^{-i\theta}$. The normalization condition
reduces then to $\int_0^{\infty}dI\,W_0(I;t)=1$ whereas the
amplitudes of other harmonics are expressed in terms of the
matrix elements $\langle n+m | \hat{\rho} | n\rangle$ of the
density operator along the $m$th collateral diagonal as
\begin{equation}\label{Harm}
\begin{array}{c}
W_m(I;t)=\frac{2}{\hbar}\,e^{-\frac{2}{\hbar}I}\sum_{n=0}^{\infty}(-1)^n
\sqrt{\frac{n!}{(n+m)!}}\times\\
\left(4I/\hbar\right)^{\frac{m}{2}}
L_n^m\left(4I/\hbar\right)\langle n+m |{\hat\rho}(t)|n\rangle,
\;\; m\ge 0,
\end{array}
\end{equation}
with $L_n^m$ Laguerre polynomials and $W_{-m}=W_m^*$. The inverted relation
reads
\begin{equation}\label{RHarm}
\begin{array}{c}
\langle n+m\big|{\hat{\rho}}(t)\big|n\rangle=(-1)^n\,
2\sqrt{\frac{n!}{(n+m)!}}\times\\
\int_0^{\infty}\!dI\,e^{-2\frac{I}{\hbar}}\,
\left(4I/\hbar\right)^{\frac{m}{2}}
L_n^m\left(4I/\hbar\right)\, W_m\left(I;t\right)\,.
\end{array}
\end{equation}

Aiming to connect the reversibility of the motion with the
complexity of the Wigner function, we follow the approach
developed in Ref.~\cite{ikeda}. We consider first the forward
evolution ${\hat\rho}(t)={\hat U(t)}{\hat\rho}(0){\hat
U}^\dagger(t)$ of a simple initial (generally mixed) state
${\hat\rho}(0)$ up to some time $t=T$. An instantaneous Hermitian
perturbation $\xi{\hat V}\,\delta(t-T)$ with the intensity
$\xi$ is then applied which transforms the state ${\hat\rho}(T)$
into ${\hat{\tilde\rho}}(T,\xi)={\hat P}(\xi){\hat\rho}(T){\hat
P}^\dagger(\xi)\,.$ The resulting transformation ${\hat P}(\xi)
=e^{-i\xi{\hat V}}$ is unitary. For example this transformation
is equivalent to the global rotation $W(I,\theta;T)\rightarrow
W(I,\theta+\xi;T)$ by the angle $\xi$ in the phase plane if the
operator ${\hat V}={\hat n}$. In particular, we use below an
infinitesimal perturbation of such a kind to reveal complexity
of the Wigner function at the instant $T$.

The new state ${\hat{\tilde\rho}}(T,\xi)$ serves as the initial
condition for the backward evolution ${\hat U}(-T)={\hat
U}^{\dag}(T)$ during the same time $T$, after which the
reversed state
\begin{equation}\label{RevState}
{\hat{\tilde\rho}}(0|T,\xi)={\hat
U}^{\dag}(T){\hat{\tilde\rho}}(T,\xi){\hat U}(T)=
{\hat P}(\xi,T){\hat\rho}(0){\hat P}^\dagger(\xi,T),
\end{equation}
is finally obtained. Here ${\hat P}(\xi,T)\equiv e^{-i\xi {\hat
V}(T)}$, with ${\hat V}(T)\equiv {\hat U}^\dagger(T){\hat
V}{\hat U}(T)$ being the Heisenberg evolution of the
perturbation during the time $T$. At last, we consider the
distance between the reversed ${\hat{\tilde\rho}}(0|T,\xi)$ and
the initial ${\hat\rho}(0)$ states, as measured by the Peres
fidelity ${\displaystyle F_{rev}(\xi;T)={\rm
Tr}[{\hat{\tilde\rho}}(0|T,\xi){\hat\rho(0)}]/ {\rm
Tr}[{\hat\rho}^2(0)]}$~\cite{peres} which can be also expressed
with the help of Eq. (\ref{RHarm}) in terms of the Wigner
function as~\cite{prosen, harmonicslong}
\begin{equation}\label{FidW}
\begin{array}{c}
{\displaystyle F_{rev}(\xi;T)=\frac{\int
d^2\alpha\,W\left(\alpha^*,\alpha;0\right)
\tilde{W}\left(\alpha^*,\alpha;0\big|T,\xi\right)}
{\int d^2\alpha\,W^2\left(\alpha^*,\alpha;0\right)}=}\\
{\displaystyle\frac{\int
d^2\alpha\,W\left(\alpha^*,\alpha;T\right)
\tilde{W}\left(\alpha^*,\alpha;T,\xi\right)}{\int
d^2\alpha\,W^2\left(\alpha^*,\alpha;T\right)}\equiv
\displaystyle F(\xi;T)\,.}
\end{array}
\end{equation}
The fidelity is bounded in the interval $[0,1]$ and is the
closer to unity the more similar are the initial and reversed
states. At the second line, which is a consequence of the
unitary time evolution, the fidelity $F(\xi;T)$ measures the
\textit{complexity} of the Wigner function at the moment $t=T$
(see below). Both the functions ${\displaystyle
F_{rev}(\xi;T)}$ and ${\displaystyle F(\xi;T)}$ are numerically
identical. The relation~(\ref{FidW}) plays the key role in our
analyses. It allows us not only to relate the degree of
reversibility to the complexity of the state at the reversal
time $T$ but also {\it to establish a strong restriction on the
upgrowth} of the number ${\cal M}(t)$ of harmonics of the
Wigner function. The crucial point is that while in classical
mechanics the number of Fourier components has no direct
physical meaning, in quantum mechanics the number of the
components of the Wigner function at any given time is related
to the \emph{degree of excitation} of the system (see for
example eq. (\ref{m_ver_n}) below) and therefore unrestricted
exponential growth of this number is not
allowed~\cite{chirikov,gu,brumer}. It is important that the
representation (\ref{FidW}) is valid not only in the quantum
but also in the classical case when the Wigner function reduces
to the classical phase-space distribution function
$W_c(\alpha^*,\alpha;t)\,$.

First of all, let us establish a connection between fidelity
and complexity of the Wigner function at an arbitrary moment
$t$. To do this we make the instant rotation $e^{-i\xi{\hat
n}}$ at this moment and utilize then the expression given in
the second line in Eq. (\ref{FidW}). This
yields~\cite{harmonicslong}
\begin{equation}\label{FidRot}
\begin{array}{c}
F(\xi;t)=1-2\sum_{m=1}^{\infty}\sin^2\left(\xi m/2\right)
\mathcal{W}_m(t)=\\ 1-\frac{1}{2}\xi^2\,\langle m^2\rangle_t+O(\xi^4)\,,
\end{array}
\end{equation}
where $\langle m^2\rangle_t=\sum_{m=0}^{+\infty} m^2\,\mathcal{W}_m(t)$
and
\begin{equation}\label{tildeW_m}
\mathcal{W}_{m\geq 0}(t)=\frac{(2-\delta_{m 0})\int_0^\infty dI
|W_{m }(I;t)|^2} {\sum_{m=-\infty}^{+\infty}
\int_0^\infty dI |W_m(I;t)|^2}\,.
\end{equation}
We define the number of developed by the time $t$ harmonics of
the Wigner function of a classically chaotic quantum system as
${\cal M}(t)= \sqrt{\langle m^2\rangle}_t$ what is in line with
the refs.~\cite{gu,brumer}. The set of positive definite
quantities $\mathcal{W}_m$ is normalized to unity,
$\sum_{m=0}^{+\infty} \mathcal{W}_m=1$, and can be therefore
given a probabilistic interpretation. Notice that our
definition of the number ${\cal M}(t)$ is applicable to
classical dynamics as well, provided that the harmonics of the
classical distribution function $W_c$ instead of those of the
Wigner function $W$ are used in Eq.~(\ref{tildeW_m}). Hence,
the number of harmonics at arbitrary time $t$ is defined via
Peres fidelity as
\begin{equation}\label{}
{\cal M}^2(t)\equiv \langle m^2\rangle_t= -\frac{d^2
F(\xi;t)}{d\xi^2}\Big|_{\xi=0}
\end{equation}

As an illustrative example we consider further the kicked
quartic oscillator defined by the Hamiltonian
~\cite{Berman78,chirikov,Sokolov84,Sokolov07}
\begin{equation}\label{Ham}
{\hat H}({\hat a}^{\dag},{\hat a})=\hbar\,\omega_0 {\hat n}+
\hbar^2\,{\hat n}^2-\sqrt{\hbar}\,g(t)({\hat a}+{\hat a}^{\dag}),
\end{equation}
where $g(t)=g_0\sum_s\delta(t-s)$. In our units, the time and
parameters $\hbar, \omega_0$ as well as the strength of the driving
force are dimensionless. The period of the driving force is set to
one. The classical dynamics of such an oscillator becomes chaotic when
the kicking strength $g_0$ exceeds a critical value $g_{0,c}\approx 1$.
The angular phase correlations decay in this case exponentially and
the mean action grows diffusively with the diffusion coefficient
$D\approx g_0^2.$

In the chosen model, the only difference between the classical
and corresponding quantum Liouville operators consists in the
substitution $\left(\omega_0+2|\alpha|^2\Rightarrow\omega_0-
\hbar-\frac{1}{2}\hbar^2\frac{\partial^2}{\partial\alpha^*
\partial\alpha}+2|\alpha|^2\right)\times\\
{\displaystyle\left(\alpha^*\partial/\partial\alpha^*-
\alpha\,\partial/\partial\alpha\right)}$ in the unperturbed
($g_0=0$) part ${\cal L}^{(0)}$. Opposite to the continuous
function $2|\alpha|^2$, the spectrum of the operator ${\hat
K}=-\frac{1}{2}\hbar^2\frac{\partial^2}
{\partial\alpha^*\partial\alpha}+2|\alpha|^2$, which coincides
formally with the Hamiltonian of a two-dimensional linear
oscillator, is discrete. This reflects the quantization of the
phase space which stays \emph{ultima analysi} behind the much
slower growth of the number of Harmonics of the Wigner function
than that of the corresponding classical phase space
distribution.

\begin{figure}[!t]
\includegraphics[width=8.0cm,angle=0]{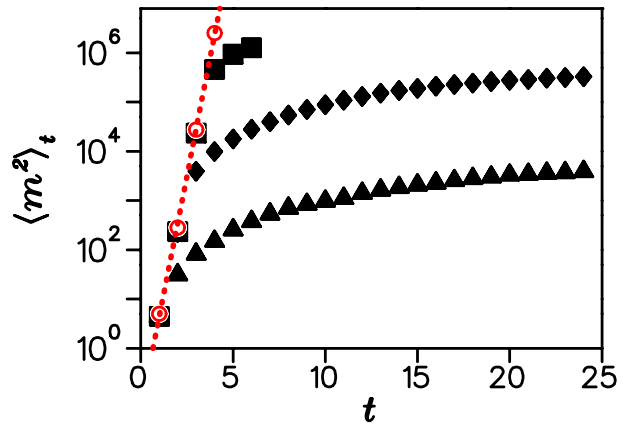}
\caption{(color online) Root-mean-square $\langle m^2\rangle_t$
versus time $t$ at $g_0=1.5$. Squares, diamonds and triangles
correspond to $\hbar = 0.01, 0.1$ and 1. Empty circles refer to
classical dynamics and the dashed line fits these data.}
\label{fig:Wquantclass}
\end{figure}

A numerical illustration of this statement is given in
Fig.~\ref{fig:Wquantclass}. The initial state is chosen to be a
pure ground state ${\hat \rho}(0)= |0\rangle\langle 0|$ which
corresponds to the isotropic Wigner function
$W(\alpha^*,\alpha;0)=2\,e^{-2|\alpha|^2}$ with the size 1/2.
This size is kept constant throughout all calculations whereas
the quantum Liouville equation is solved for a number of
decreasing values of the effective Planck's constant $\hbar$
thus approaching the classical dynamics. Let us notice that for
$\hbar<1$ our initial condition corresponds to an incoherent
mixture of eigenstates of the operator ${\hat H}^{(0)}$. It is
clearly seen that the exponential increase of $\langle m^2
\rangle_t$ takes place only up to the Ehrenfest time scale $t_E
\propto \ln \hbar$~\cite{Berman78}, consistently with the
findings reported in Refs.~\cite{gu,brumer}. Then, a much
slower power-law increase follows.

To explain such a behavior we turn to the equivalent
representation of fidelity given in the first line of
Eq.~(\ref{FidW}). We restrict ourselves to the case of the pure
ground initial state. More general consideration which includes
also arbitrary incoherent initial mixtures is presented
in~\cite{harmonicslong}. Using this representation we find in
this case that $F_{rev}(\xi,t)=|f(\xi,t)|^2$
where~\cite{harmonicslong}
\begin{equation}\label{Fid_n}
\begin{array}{c}
f(\xi,t)=\langle 0|e^{-i\xi{\hat n}(t)}|0\rangle=
{\rm Tr}\left[e^{-i\xi{\hat n}}\rho(t)\right]=\\
\frac{e^{i\xi/2}}{\cos(\xi/2)}\int_0^{\infty}\,
e^{-2i\tan(\xi/2)\frac{I}{\hbar}}\,W_0(I;t)\,.
\end{array}
\end{equation}
In such a way we relate the behavior of fidelity to evolution of the action
variable. Comparing now the $\xi^2$ terms in the expansions of the both
possible representations~(\ref{FidRot}) and~(\ref{Fid_n}) of fidelity we
arrive at the following {\it significant  exact relation} between the number
of harmonics and the root-mean-square deviation of the action
\begin{equation}\label{m_ver_n}
\langle m^2\rangle_t=2\,\chi_2(t), \,\,\,\,\,
\chi_2(t)\equiv\frac{1}{\hbar^2}\left(\langle I^2\rangle_t-
\langle I\rangle^2_t\right)\,.
\end{equation}

\begin{figure}[!t]
\includegraphics[height=7.cm,angle=90]{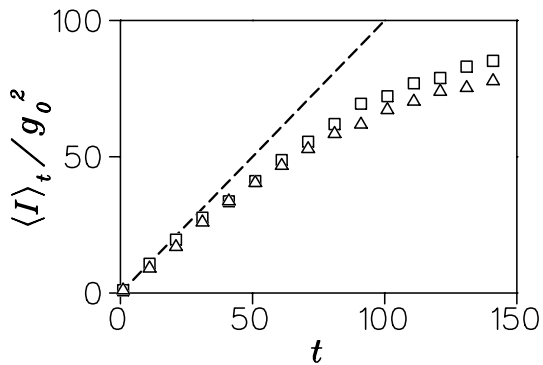}
\caption{Mean value $\langle I\rangle_t/ g_0^2$ as a function
of time $t$. Squares and triangles correspond to
$(\hbar,g_0)$=$(1,2)$ and $(2,3)$. The straight line shows the
classical diffusion law $\langle I\rangle_t=~g_0^2 t$.}
\label{fig:avn}
\end{figure}

\begin{figure}[!t]
\includegraphics[width=4.7cm,angle=90]{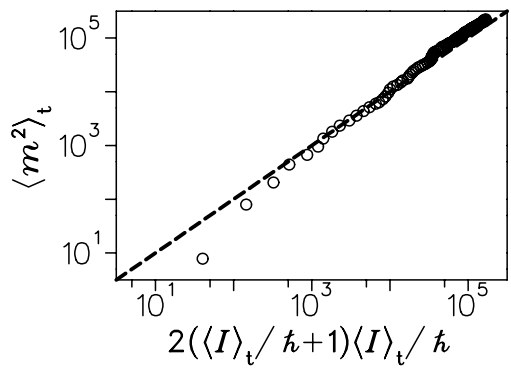}
\caption{$\langle m^2\rangle_t$ vs
$\chi_2(t)\approx\frac{\langle I\rangle_t}{\hbar}
\left(\frac{\langle I\rangle_t}{\hbar}+1\right)$, for
$t=1,\ldots,100$. Data correspond to $\hbar=1$ and $g_0=2$.}
\label{fig:m^2_<I>^2}
\end{figure}

Thorough numerical study~\cite{harmonicslong} convince us that
after proper averaging over strong irregular fluctuations
(\emph{coarse graining}) the zero harmonic amplitude decays
exponentially with the action $I$ at any given time $t>t_E$,
\begin{equation}\label{CG_W}
\overline{W_0(I;t)}=\frac{1}{\langle I\rangle_t+\frac{\hbar}{2}}
\exp\left(-\frac{I}{\langle
I\rangle_t+\frac{\hbar}{2}}\right)\,.
\end{equation}
It follows from Eq. (\ref{CG_W}) that
$\chi_2(t)\approx\frac{\langle I\rangle_t}{\hbar}
\left(\frac{\langle I\rangle_t}{\hbar}+1\right)$ and,
therefore, ${\cal M}(t)\approx\sqrt{2}\frac{\langle
I\rangle_t}{\hbar}$. The time dependence of the mean action
$\langle I\rangle_t$ (the deterministic quantum diffusion) is
shown in Fig.~\ref{fig:avn} whereas the validity of the stated
connection between the number of the harmonics and the
excitation of the system is illustrated in
Fig~\ref{fig:m^2_<I>^2}. Thus the number of harmonics grows
after the Ehrenfest time not faster than linearly.

\begin{figure}
\includegraphics[width=7.cm,angle=0]{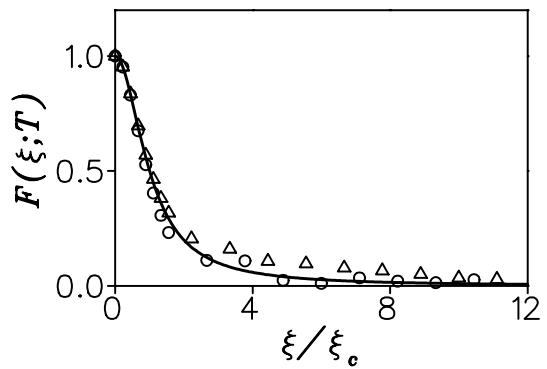}
\caption{Fidelity $F(\xi;T)$ versus the scaled variable
$\xi/{\xi}_c(T)$. Data correspond to: (a) $\hbar = 1, g_0 = 2$;
circles: $T = 10$; triangles: $T =
50$. The full curve shows the theoretical prediction
Eq.~(\ref{Fid_m^2}). The deviations on the tail of the curve
are due to fluctuations neglected in~(\ref{Fid_m^2}).}
\label{fig:fig3}
\end{figure}

It can be readily shown now that for any finite $\xi\ll 1$
fidelity equals in the approximation~(\ref{CG_W})
\begin{equation}\label{Fid_m^2}
F_{rev}(\xi;T)=F(\xi;t=T)=\frac{1}{1+\frac{1}{2}\xi^2\langle m^2\rangle_T}\,.
\end{equation}
More than that, this formula is valid for any time including
the times $T<T_E$~\cite{harmonicslong}. The found result shows
that a crossover takes place near the {\it critical value}
$\xi_c(T)\equiv\sqrt{2}/{\cal M}(T)$, from good,
$F(\xi;T)\approx 1$, to broken, $F(\xi;t)\approx
(\xi/\xi_c(T))^2\ll 1$, reversibility. Our numerical
simulations~(see Fig.~\ref{fig:fig3})
nicely confirm the formula~(\ref{Fid_m^2}).

The dependence of $\xi_c(T)$ on the reversal time $T$ is
strikingly different within and outside the Ehrenfest time
scale. When in the former case it drops exponentially in
accordance with the classical exponential proliferation of the
number of harmonics, in the latter case it decreases at most
linearly. This explains the numerically discovered~\cite{arrow}
much weaker sensitivity of quantum dynamics to perturbations as
compared to classical dynamics.

To summarize, we have established a quantitative relation
between complexity of the Wigner function and degree of
reversibility of motion of a classically chaotic quantum
system. We have analytically proved that the number of
harmonics ${\cal M}(t)$ of this function, which can serve as a
natural measure of the complexity, increases after the
Ehrenfest time \textit{not faster than linearly} in striking
contrast with classical dynamics, where the number of harmonics
of phase space distribution growths exponentially. We have
shown that if a quantum motion has been perturbed at some
moment $T$ by an external force with intensity $\xi$ and then
reversed, its initial state is recovered with the accuracy
$\sim(\xi/\xi_c)^2$ as long as the strength is restricted to
the interval $0<\xi<\xi_c(T)= \sqrt{2}/{\cal M}(T)$. This
interval decreases with the time $T$ at most linearly beyond
the semi-classical domain but shrinks exponentially due to the
classical exponential instability when the this domain is
approached.

The above outlined phase-space approach is quite general and can be
readily extended to systems with arbitrary number of degrees of freedom,
including qubit systems, whose Hamiltonian can be expressed in terms of a
set of bosonic creation-annihilation operators. Therefore, this approach
could shed some light of the connection between complexity and entanglement,
a fundamental issue of great relevance for the prospects of quantum information science~\cite{Benenti08}.

{\it Acknowledgements.}
It is our pleasure to thank Giuliano Benenti for fruitful collaboration and
illuminating discussions. We also are very indebted to Giulio Casati for
drawing our attention to the problem of quantum reversibility. We acknowledge
support by the Cariplo Foundation, INFN and by the RAS Joint scientific program "Nonlinear dynamics and Solitons".


\begin{thebibliography}{01}
\bibitem{fidelitynote}
We note that, similarly to Ref.~\cite{heller} and in contrast
to other previous studies of the fidelity~\cite{prosen},
the backward evolution proceeds with the same Hamiltonian as the
forward evolution and the perturbation acts instantaneously
only at the reversal time $T$.
\end{thebibliography}
\end{document}